\documentclass[aps,prd,showpacs,superscriptaddress,twocolumn,10pt]{revtex4}

\usepackage{color,amsmath,amssymb,graphicx,latexsym,txfonts}
\usepackage{threeparttable,multirow,subfigure,booktabs,lineno,tabularx,ulem}

\newcommand{\vecA}{\boldsymbol{A}}
\newcommand{\vecx}{\boldsymbol{x}}

\begin{document}


\title{High-precision search for dark photon dark matter with the Parkes Pulsar Timing Array}

\author{Xiao Xue}
\affiliation{II. Institut f\"{u}r Theoretische Physik, Universit\"at Hamburg, Luruper Chaussee 149, D-22761 Hamburg}

\author{Zi-Qing Xia}
\affiliation{Key Laboratory of Dark Matter and Space Astronomy, Purple Mountain Observatory, Chinese Academy of Sciences, Nanjing 210023, China}

\author{Xingjiang Zhu}
\affiliation{Advanced Institute of Natural Sciences, Beijing Normal University at Zhuhai 519087, China}
\affiliation{School of Physics and Astronomy, Monash University, Clayton, VIC 3800, Australia}
\affiliation{OzGrav: The ARC Centre of Excellence for Gravitational Wave Discovery, Hawthorn, VIC 3122, Australia}
 
\author{Yue Zhao}
\affiliation{Department of Physics and Astronomy, University of Utah, Salt Lake City, UT 84112, USA}

\author{Jing Shu\footnote{jshu@itp.ac.cn}}
\affiliation{CAS Key Laboratory of Theoretical Physics, Institute of Theoretical Physics, Chinese Academy of Sciences, Beijing 100190, China}
\affiliation{School of Physical Sciences, University of Chinese Academy of Sciences, Beijing 100049, China}
\affiliation{CAS Center for Excellence in Particle Physics, Beijing 100049, China}
\affiliation{Center for High Energy Physics, Peking University, Beijing 100871, China}
\affiliation{School of Fundamental Physics and Mathematical Sciences, Hangzhou Institute for Advanced Study, University of Chinese Academy of Sciences, Hangzhou 310024, China}
\affiliation{International Centre for Theoretical Physics Asia-Pacific, Beijing/Hangzhou, China}

\author{Qiang Yuan\footnote{yuanq@pmo.ac.cn}}
\affiliation{Key Laboratory of Dark Matter and Space Astronomy, Purple Mountain Observatory, Chinese Academy of Sciences, Nanjing 210023, China}
\affiliation{Center for High Energy Physics, Peking University, Beijing 100871, China}
\affiliation{School of Astronomy and Space Science, University of Science and Technology of China, Hefei 230026, China}

\author{N. D. Ramesh Bhat}
\affiliation{International Centre for Radio Astronomy Research, Curtin University, Bentley, WA 6102, Australia}

\author{Andrew D. Cameron}
\affiliation{OzGrav: The ARC Centre of Excellence for Gravitational Wave Discovery, Hawthorn, VIC 3122, Australia}
\affiliation{Australia Telescope National Facility, CSIRO Astronomy and Space Science, P.O. Box 76, Epping, NSW 1710, Australia}
\affiliation{Centre for Astrophysics and Supercomputing, Swinburne University of Technology, P.O. Box 218, Hawthorn, VIC 3122, Australia}

\author{Shi Dai}
\affiliation{School of Science, Western Sydney University, Locked Bag 1797, Penrith South DC, NSW 2751, Australia}
\affiliation{Australia Telescope National Facility, CSIRO Astronomy and Space Science, P.O. Box 76, Epping, NSW 1710, Australia}

\author{Yi Feng}
\affiliation{Research Institute of Artificial Intelligence, Zhejiang Lab, Hangzhou, Zhejiang 311121, China}

\author{Boris Goncharov}
\affiliation{School of Physics and Astronomy, Monash University, Clayton, VIC 3800, Australia}
\affiliation{OzGrav: The ARC Centre of Excellence for Gravitational Wave Discovery, Hawthorn, VIC 3122, Australia}

\author{George Hobbs}
\affiliation{Australia Telescope National Facility, CSIRO Astronomy and Space Science, P.O. Box 76, Epping, NSW 1710, Australia}

\author{Eric Howard}
\affiliation{Australia Telescope National Facility, CSIRO Astronomy and Space Science, P.O. Box 76, Epping, NSW 1710, Australia}
\affiliation{Macquarie University, Department of Physics and Astronomy, Sydney, NSW, 2109, Australia}

\author{Richard N. Manchester}
\affiliation{Australia Telescope National Facility, CSIRO Astronomy and Space Science, P.O. Box 76, Epping, NSW 1710, Australia}

\author{Aditya Parthasarathy}
\affiliation{Centre for Astrophysics and Supercomputing, Swinburne University of Technology, P.O. Box 218, Hawthorn, VIC 3122, Australia}
\affiliation{Max-Planck-Institut f\"{u}r Radioastronomie, Auf dem H\"{u}gel 69, D-53121 Bonn, Germany}

\author{Daniel J. Reardon}
\affiliation{OzGrav: The ARC Centre of Excellence for Gravitational Wave Discovery, Hawthorn, VIC 3122, Australia}
\affiliation{Centre for Astrophysics and Supercomputing, Swinburne University of Technology, P.O. Box 218, Hawthorn, VIC 3122, Australia}

\author{Christopher J. Russell}
\affiliation{CSIRO Scientific Computing, Australian Technology Park, Locked Bag 9013, Alexandria, NSW 1435, Australia}

\author{Ryan M. Shannon}
\affiliation{OzGrav: The ARC Centre of Excellence for Gravitational Wave Discovery, Hawthorn, VIC 3122, Australia}
\affiliation{Centre for Astrophysics and Supercomputing, Swinburne University of Technology, P.O. Box 218, Hawthorn, VIC 3122, Australia}

\author{Ren\'ee Spiewak}
\affiliation{Centre for Astrophysics and Supercomputing, Swinburne University of Technology, P.O. Box 218, Hawthorn, VIC 3122, Australia}
\affiliation{Jodrell Bank Centre for Astrophysics, University of Manchester, Manchester M13 9PL, UK}

\author{Nithyanandan Thyagarajan}
\affiliation{National Radio Astronomy Observatory, 1003 Lopezville Rd, Socorro, NM 87801, USA}

\author{Jingbo Wang}
\affiliation{Xinjiang Astronomical Observatory, Chinese Academy of Sciences, 150 Science 1-Street, Urumqi, Xinjiang 830011, China}

\author{Lei Zhang}
\affiliation{School of Physics and Technology, Wuhan University, Wuhan 430072, China}

\author{Songbo Zhang}
\affiliation{Purple Mountain Observatory, Chinese Academy of Sciences, Nanjing 210023, China}

\collaboration{PPTA Collaboration}

\date{\today}

\begin{abstract}
The nature of dark matter remains obscure in spite of decades of 
experimental efforts. The mass of dark matter candidates can span 
a wide range, and its coupling with the Standard Model sector 
remains uncertain. All these unknowns make the detection of dark 
matter extremely challenging. Ultralight dark matter, with 
$m \sim10^{-22}$ eV, is proposed to reconcile the disagreements 
between observations and predictions from simulations of small-scale 
structures in the cold dark matter paradigm, while remaining 
consistent with other observations. Because of its large de Broglie 
wavelength and large local occupation number within galaxies, 
ultralight dark matter behaves like a coherently oscillating 
background field with an oscillating frequency dependent on its mass. 
If the dark matter particle is a spin-1 dark photon, such as the 
$U(1)_B$ or $U(1)_{B-L}$ gauge boson, it can induce an external 
oscillating force and lead to displacements of test masses. 
Such an effect would be observable in the form of periodic 
variations in the arrival times of radio pulses from highly stable 
millisecond pulsars. In this study, we search for evidence of 
ultralight dark photon dark matter (DPDM) using 14-year high-precision 
observations of 26 pulsars collected with the Parkes Pulsar Timing 
Array. While no statistically significant signal is found, we place 
constraints on coupling constants for the $U(1)_B$ and $U(1)_{B-L}$ 
DPDM. Compared with other experiments, the limits 
on the dimensionless coupling constant $\epsilon$ achieved in our 
study are improved by up to two orders of magnitude when the dark 
photon mass is smaller than $3\times10^{-22}$~eV ($10^{-22}$~eV) 
for the $U(1)_{B}$ ($U(1)_{B-L}$) scenario.
\end{abstract}

\maketitle


{\it Introduction.} ---
About 26$\%$ of the total energy in our present-day Universe is composed 
of dark matter. The cold dark matter paradigm has been widely accepted 
since it explains most of the cosmological observations over a large span 
of redshift \cite{Bahcall:1999xn}. However, numerical simulations of the 
traditional particle like cold dark matter models show that the central 
density profile of dark matter in dwarf galaxies is much steeper than 
that inferred from observed rotation curves, and the number of satellite 
galaxies of Milky Way-like hosts predicted from simulations is higher 
than that inferred from observations~\cite{Weinberg2015}. These so-called 
``core-cusp problem'' and ``missing-satellite problem'' impose challenges 
to the the conventional cold dark matter paradigm. Baryonic feedback effects \cite{chan2015impact} may be a viable solution, but 
it is a nontrivial task. Meanwhile, various alternative dark matter 
models have been proposed to address such small-scale shortcomings, 
e.g., warm dark matter \cite{bode2001halo}, self-interacting dark 
matter~\cite{Tulin:2017ara}, and fuzzy dark matter \cite{Hu2000}. 

A dark photon is a hypothetical particle from the theory beyond the 
Standard Model of particle physics. Just like the ordinary photon, 
a dark photon is the gauge boson of a $U(1)$ interaction. The existence 
of the dark photon is well predicted in many string-inspired models, 
such as large-volume string compactifications 
\cite{Burgess:2008ri,Goodsell:2009xc,Cicoli:2011yh}. 
The dark photon mass can be generated by either the Higgs mechanism 
or the Stueckelberg mechanism, and it is naturally light \cite{fn1}. 
When this $U(1)$ symmetry is the baryon or lepton number or their 
linear combination, protons, neutrons, and electrons are ``charged'' 
under this symmetry, and there will be an extra force (i.e., the 
``fifth force") between objects that consist of ordinary matter.

In recent years, the dark photon has been widely considered as a viable 
dark matter candidate. Interestingly, when the dark photon is ultralight 
($m_A\sim10^{-22}$~eV), which represents a realization of the fuzzy dark 
matter paradigm, its de Broglie wavelength is up to subgalactic scale, 
i.e., $O(0.1-1)$ kpc, and the occupation number in dark matter halos is 
very large~\cite{fn7}. These aspects imply that the ``wave nature'' of dark 
matter particles is pronounced; hence the dark matter can be properly 
treated as an oscillating background field rather than individual particles. 
Because of the formation of a ``soliton core'' instead of a density cusp 
at the center of galaxies \cite{Schive2014a,Schive2014b,Zhang2018}, 
the density of the dark matter halo at a galactic center becomes flat. 
This provides a better fit to observations than a cuspy Navarro-Frenk-White 
profile \cite{NFW1997} predicted in the cold dark matter scenario. 
In addition, the substructures formed under the fuzzy dark matter 
hypothesis are fragile against tidal disruptions, since they usually 
have less concentrated mass distributions. This solves the 
``missing-satellite'' problem \cite{Hui2017} at the same time. 
To date, the ultralight dark photon dark matter (DPDM) leads to predictions 
consistent with most existing observations at various scales \cite{Hui2017}, 
making it one of the most compelling dark matter candidates.

Conventional direct detection experiments \cite{PANDAX2017,LUX2017,XENON2018} 
are not sensitive to fuzzy dark matter particles because of the extremely 
small energy and momentum depositions in elastic scatterings. If the 
ultralight dark matter particle is a $U(1)_B$ or $U(1)_{B-L}$ gauge boson 
(i.e., dark photon), the fifth-force experiments can be used to put 
stringent constraints on the existence of such particle \cite{MICROSCOPE}, as well as the energy loss of compact binary systems \cite{Binary1}. Note that, a more general $U(1)$ interaction that includes both $U(1)_B$ and $U(1)_{B-L}$ is discussed in earlier work \cite{PierreTheory3,PierreTheory4}, while the fifth-force constraint is derived in Ref.~\cite{Pierre1}. Furthermore, the dark matter background can cause displacements on 
terrestrial/celestial objects, leading to observable effects \cite{Graham2015}.
For example, depending on the mass of dark photon particles which determines 
the dark matter oscillation frequency, such motion can be probed using 
high-precision astrometry \cite{Guo2019}, spectroscopic \cite{Armengaud2017}, 
timing observations \cite{DeMartino2017,Porayko2018}, as well as 
gravitational-wave detectors \cite{Pierce2018,Guo:2019ker, LIGOScientific:2021odm}. 

Among all these relevant observations, the pulsar timing array (PTA) 
experiments are of particular interest to us. Millisecond pulsars with very 
high rotational stability are observed to emit periodic electromagnetic 
pulses with incredible accuracy. For the best pulsars, the measurement 
uncertainties of pulse arriving times are below $100$ ns. The information 
of possible new physics (including signatures of DPDM) is in the 
irregularities of pulse arrival times, which are usually called timing 
residuals. The $U(1)_B$ or $U(1)_{B-L}$ DPDM with an ultralight mass 
$ \sim 10^{-22} {~\rm eV}$, induces displacements 
of Earth and pulsar that cause a periodic signal in timing residuals with 
frequency \cite{fn2} $f=m_A/2\pi\approx 2.4\times 10^{-8}{~\rm Hz}$. This frequency 
falls into the sensitive region of PTA experiments, which 
have collected measurements of pulse arrival times on weekly to monthly 
cadence over time scales of 10–20 years.

In this work, we make use of the second public data release of the Parkes
PTA project \cite{PPTA,Kerr2020parkes}, to search for evidence of the 
DPDM. Subsets of these data have been used to search for stochastic 
gravitational waves \cite{Shannon2015} as well as the gravitational 
effects induced by ultralight scalar dark matter \cite{Porayko2018}.
Compared with Ref.~\cite{Porayko2018}, we discuss the direct 
coupling between DPDM and the ordinary matter in this work, which 
results in different signals from the gravitational effect studied in Ref.~\cite{Porayko2018}. We also improve the analysis through
considering the correlation of pulsars in the DPDM field, the
interference of DPDM wave functions, and excluding fake signals that arise from the time-frequency method. (see Secs. I and II of
the Supplemental Material \cite{supp} for details.)

{\it Observations.} --- The second data release 
(DR2 \cite{fn3}) of the PPTA 
project \cite{Kerr2020parkes} includes high-precision timing observations 
for 26 pulsars collected with the 64-m Parkes radio telescope in Australia. 
The data span is 14.2 years, from 2004 February 6 to 2018 April 25, 
except for PSR J0437$-$4715 where pre-PPTA observations from 2003 April 
12 are also included. The end date of this data set corresponds to the 
installation and commissioning of a new ultrawide bandwidth receiver 
on the Parkes telescope. The observations were made typically once every 
3 weeks in three radio bands (10, 20, and 40/50 cm). 
Details of the observing systems and data processing procedures are 
described in Refs.~\cite{PPTA,Kerr2020parkes}.

Pulsar times of arrival (ToAs) in PPTA DR2 are obtained using the Jet 
Propulsion Laboratory Solar-System ephemeris DE436 and the TT(BIPM18) 
reference time scale published by the International Bureau of Weights 
and Measures (BIPM). We fit these ToAs to a timing model with the standard 
{\tt TEMPO2} \cite{software_tempo2_1,software_tempo2_2} software package.
We perform the Bayesian noise analysis and search for DPDM signals 
using the {\tt enterprise} \cite{fn4} 
package. The {\tt PTMCMC} \cite{fn5}
sampler is used for the stochastic sampling of parameter space and the 
calculation of Bayesian upper limits. We also use 
{\tt PyMultiNest} \cite{fn6}
to calculate the Bayes Factor while performing model selection. 
The names of pulsars, basic observing information, and noise properties 
of the PPTA DR2 pulsars are listed in Supplemental Table~S1 \cite{supp}.

{\it Timing model.} --- The ToAs of a pulsar include quite a few terms.
There are several astronomical effects that should be accounted for, 
such as the intrinsic pulsar spin-down, the proper motion of the pulsar, 
the time delay due to interstellar medium, and the Shapiro delay caused 
by Solar-System planets. In addition to these deterministic effects, 
both uncorrelated noise (white noise) and correlated noise (red noise) 
may be present in pulsar timing data \cite{red_noise_1,red_noise_2,
red_noise_exp_1,red_noise_exp_2,DM_noise_1,DM_noise_2}. 
We model the ToAs as follows:
\begin{equation}
t = t_{\rm TM}(\boldsymbol{\varphi} _{\rm TM}) + {\Delta t}_{\rm Noise}(\boldsymbol{\vartheta}_{\rm Noise}) + {\Delta t}_{\rm DPDM}(\boldsymbol{\vartheta}_{\rm DPDM}),
\end{equation}
where $t_{\rm TM}$ is the ``timing model'' that accounts for the 
deterministic effects, ${\Delta t}_{\rm Noise}$ represents stochastic 
noise terms, and $\boldsymbol{\varphi}_{\rm TM}$ and 
$\boldsymbol{\vartheta}_{\rm Noise}$ are parameters in the timing model 
and noise model, respectively. ${\Delta t}_{\rm DPDM}$ is used to 
describe the ``signal'' term caused by the DPDM with paramaters
$\boldsymbol{\vartheta}_{\rm DPDM}$. We make use the 
{\tt TEMPO2} and {\tt enterprise} software packages to determine the 
timing model and marginalize over model uncertainties in our Bayesian 
analysis. More details can be found in Refs.~\cite{Porayko2018,temponest}.

The noise term includes white noise and red noises from the rotational
irregularities of the pulsar, the variations of the dispersion measure, 
and the band noise. The noise is assumed to follow a multivariate 
Gaussian distribution with a covariance matrix. For the red noises,
power-law frequency dependence is assumed with parameters independently
fitted for each noise term. We describe the modelings of noise in 
detail in the Supplemental Material \cite{supp}.

{\it The DPDM-induced ToA residuals.} ---
Within coherence region where  $l < l_c \sim 2\pi/(m_A v_{\rm vir})$, 
the DPDM can be approximately described as a plane wave, 
$\boldsymbol{A}(t,\boldsymbol{x})= \boldsymbol{A_0}\sin\big(m_A t - 
\boldsymbol{k}\cdot \boldsymbol{x}+\boldsymbol{\alpha}(\boldsymbol{x})\big)$. Here $m_A$ is the dark 
photon mass, the phase term $\boldsymbol{\alpha}$ is a function of $\boldsymbol{x}$ \cite{Schive2014a,Schive2014b}, and $\boldsymbol{k}$ is the 
characteristic momentum. The direction of $\boldsymbol{k}$ is random and its magnitude is 
$\sim m_A v_{\rm vir}$ where $v_{\rm vir}$ is the virial velocity in 
our Galaxy. $\boldsymbol{A_0}$ is the gauge potential of the DPDM 
background, whose direction is another random vector and is unrelated to 
that of $\boldsymbol{k}$ in the non-relativistic limit. The averaged 
value of the magnitude $|\boldsymbol{A_0}|^2$ is determined by the 
local dark matter energy density, $2\rho_{0}/m_A^2$, with 
$\rho_{0}=0.4$~GeV~cm$^{-3}$ near the Earth \cite{Catena2010}. 
The interference among dark photons induces a random $O(1)$ 
fluctuation to the magnitude. This plane-wave approximation breaks 
down when the measurements are performed at a time scale longer than 
the coherence time, $t_c \sim 2\pi/(m_A v_{\rm vir}^2)$, or at a 
length scale larger than the coherence length $l_c$.  Note that we 
only consider the vector components of the gauge potential for 
dark photon field because we adopt the Lorentz gauge where the 
contribution from the scalar component is negligible \cite{Pierce2018}.
In the Supplemental Material \cite{supp} we describe the simulations of
the DPDM distribution in the local vicinity of the Solar System.

The weak coupling between the dark photon background and the ``dark charge'' 
results in an acceleration of a test body \cite{Pierce2018} as
\begin{equation}
\boldsymbol{a}(t,\boldsymbol{x})\simeq \epsilon e 
\frac{q}{m}m_A\boldsymbol{A_0}\cos\big(m_At-\boldsymbol{k \cdot x}+\boldsymbol{\alpha}(\boldsymbol{x})\big),
\end{equation}
where $m$ is the mass of the test body, an $\epsilon$ characterizes the 
coupling strength of the new gauge interaction that is normalized to 
the electromagnetic coupling constant $e$. The dark charge $q$ equals 
the baryon number for the $U(1)_{B}$ interaction or baryon-minus-lepton 
number for the $U(1)_{B-L}$ interaction. Such an acceleration allows 
the detection of the DPDM in several ways, e.g., using high-precision astrometry \cite{Guo2019} or gravitational-wave 
detectors \cite{Pierce2018}. 

The acceleration causes a displacement to a test object, which is approximately
\begin{equation}
\Delta\boldsymbol{x}(t,\boldsymbol{x})= - \frac{\epsilon eq}{mm_A}
\boldsymbol{A_0}\cos\big(m_At-\boldsymbol{k \cdot x}+\boldsymbol{\alpha}(\boldsymbol{x})\big).
\label{displacement}
\end{equation}
Note that here we neglect the contribution to the displacement from the spatial part $-\boldsymbol{k}\cdot \boldsymbol{x}+\boldsymbol{\alpha}(\boldsymbol{x})$. The virial velocity of dark matter is $v_{vir}\sim10^{-3} c$. For the dark photon mass range of interest, $10^{-23.5}$~eV$\leq m_A\leq 10^{-21}$~eV, the coherence length ranges from 0.04 to 13 kpc. For an observation of $O(10)$~years, the proper motion of a celestial body (the Sun or the pulsar) is about $3\times10^{-3}$~pc which is much smaller than the coherence length of the DPDM field, assuming a proper motion velocity of $\sim 10^{-3}c$. The phase change for the Earth and pulsars, induced by the $-\boldsymbol{k}\cdot \boldsymbol{x}+\boldsymbol{\alpha}(\boldsymbol{x})$ term, can be safely ignored and the DPDM field can be treated as spatially uniform around the Earth or the pulsar. 
Therefore we can rewrite Eq.~(\ref{displacement}) as
\begin{equation}\label{displacement2}
\Delta\boldsymbol{x}_{e,p}(t)= - \frac{\epsilon eq}{mm_A}
\boldsymbol{A_{0}}^{e,p}\cos(m_A t+\boldsymbol{\alpha}^{e,p}),
\end{equation}
here $\boldsymbol{A_0}^{e,p}$ and $\boldsymbol{\alpha}^{e,p}$ represent the amplitude and phase of the dark photon field at the locations of the Earth and pulsars, respectively.

Most of the pulsars studied in this work have distances between 0.1 and several kpc, which are comparable to the coherence length. Therefore, we perform the analysis in two limits:
\begin{itemize}
\item[(A)] Completely uncorrelated: The phases and amplitudes of dark photon background for each pulsar are independent.
\item[(B)] Completely correlated: The phases are independent phases but with a common amplitude.
\end{itemize}
In both cases, we treat the phases of the DPDM field at each pulsar as independent free parameters because the locations to most of these pulsars are highly uncertain.
Such an uncertainty results in a random phase to dark photon field characteristic to each pulsar. 
Note that for $m_A>10^{-22}$~eV we only perform the uncorrelated analysis (A), since in this high-mass range most of the pulsars should lie in the uncorrelated regime (see Supplemental Fig.~S2 \cite{supp}).
A hybrid of correlated and uncorrelated treatment can in principle be applied based on the comparison between the coherence length and the separation of each pulsar pair.
We expect the results of such a hybrid analysis to fall between the two cases discussed here.

We further assume a flat spacetime and obtain the time residual, $\Delta t_{\rm DPDM}$, induced by the DPDM
\begin{eqnarray}\label{delay}
\Delta t_{\rm DPDM} &=& \left(|\boldsymbol{d}+\Delta \boldsymbol{x}_p
(t^\prime)-\Delta \boldsymbol{x}_e(t)|-|\boldsymbol{d}|\right)\\\nonumber
      &\simeq& \boldsymbol{n}\cdot \left(\Delta  \boldsymbol{x}_p(t^\prime)
      -\Delta \boldsymbol{x}_e(t)\right),
\end{eqnarray}
where $t^\prime$ and $t$ are the times when the pulsar emits and the Earth receives a pulse respectively, and $\boldsymbol{d}$ is the position vector pointing from the Earth to the pulsar, and $\boldsymbol{n}=\boldsymbol{d}/|\boldsymbol{d}|$. The timing residual caused by the DPDM is anisotropic. Specifically, the Earth term is a dipole contribution in terms of angular correlation, which is distinct from the monopole signal that derives from the gravitational effect of the general fuzzy dark matter model \cite{Porayko2018,gw_effect_1}. 

Now we write the timing residuals explicitly as follows
\begin{eqnarray}
\Delta t _{\rm DPDM}^{(B)} &=&  -\frac{\epsilon e}{m_A}\left( \frac{q^{(B)}_p}{m_p}\boldsymbol{A_0}^p \cos\left(m_A t+\boldsymbol{\alpha}_p\right)\right. - \nonumber \\
& &\left.\frac{q^{(B)}_e}{m_e}\boldsymbol{A_0}^e \cos\left(m_A t+\boldsymbol{\alpha}_e\right)  \right)\cdot \boldsymbol{n}, \\
\Delta t _{\rm DPDM}^{(B-L)} &=&  -\frac{\epsilon e}{m_A}\left( \frac{q^{(B-L)}_p}{m_p}\boldsymbol{A_0}^p \cos\left(m_A t+\boldsymbol{\alpha}_p\right)\right. - \nonumber \\
& & \left.\frac{q^{(B-L)}_e}{m_e}\boldsymbol{A_0}^e \cos\left(m_A t+\boldsymbol{\alpha}_e\right)  \right)\cdot \boldsymbol{n},
\end{eqnarray}
where $q^{(B)}_{e,p}$, $q^{(B-L)}_{e,p}$, $m_{e}$ and $m_{p}$ are the $B$ charge,  $B-L$ charge, the mass of the Earth and the pulsar, respectively. For $U(1)_B$, $q^{(B)}/m$ is approximately $1/{\rm GeV}$ for both the Earth and the pulsar. For $U(1)_{B-L}$, $q^{(B-L)}/m$ is about $1/{\rm GeV}$ for the pulsar and $0.5/{\rm GeV}$ for the Earth. Note that we absorb the time difference between a pulsar and the Earth into the phase term.

\begin{figure*}
\centering
\includegraphics[width=\textwidth]{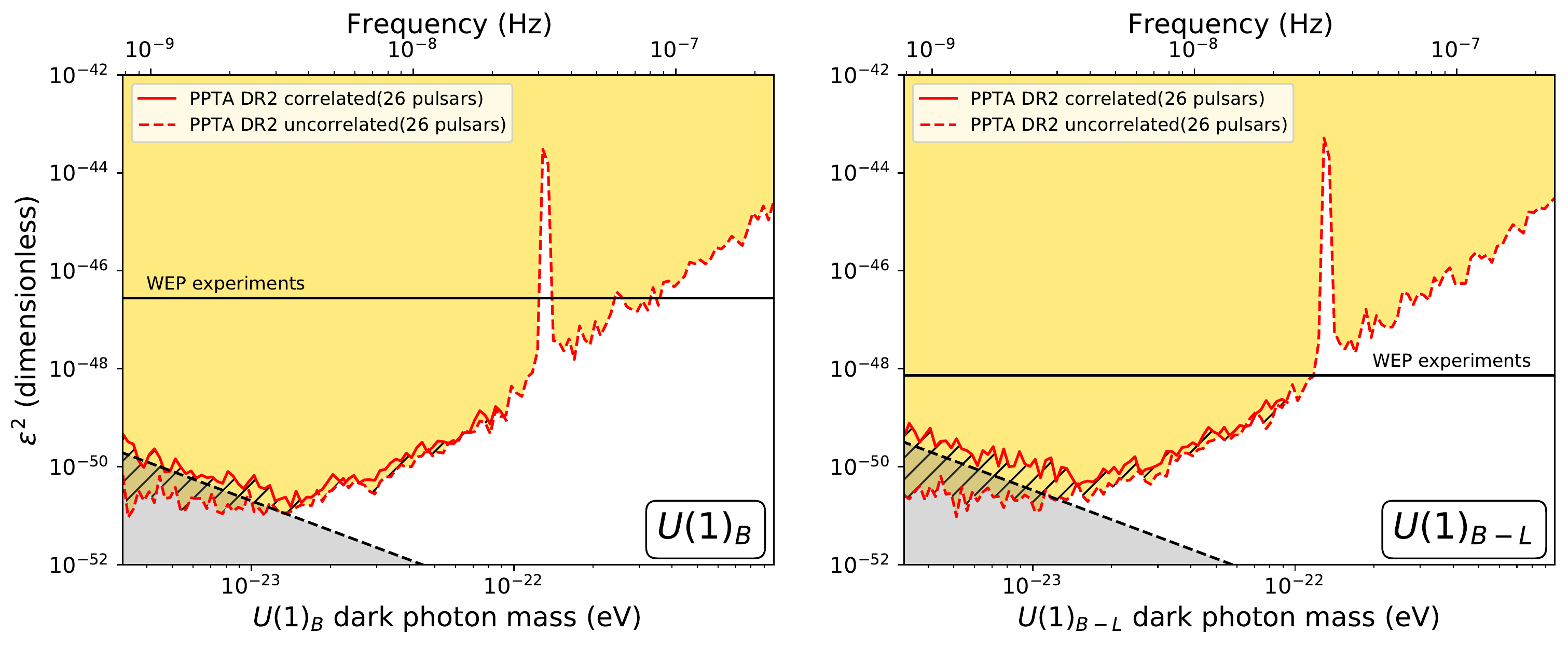}
\caption{Constraints on the dark photon mass $m_A$ and the coupling 
constant $\epsilon^2$ for the $U(1)_B$ (left) and $U(1)_{B-L}$ (right) 
gauge groups. The red solid and dashed lines are derived in this work 
using the PPTA DR2 data, under the assumptions that the dark photon 
field polarizations are correlated and uncorrelated among various pulsars, 
respectively. The horizontal solid line is the limit from the MICROSCOPE WEP experiment \cite{MICROSCOPE}. 
The black dashed line and the gray shaded region indicate the 
parameter space where the gravitational effect (studied in 
Refs.~\cite{Porayko2018,gw_effect_1,gw_effect_2}) dominates over the 
gauge interaction (studied here), for pulsar timing observations. 
More details about the WEP experiments and gravitational effects 
are presented in the Supplemental Material \cite{supp}. Note that in the WEP 
experiments, the dark photon is not required to be the dark matter.}
\label{Fig-bayes-upperlimit}
\end{figure*}

{\it Statistical analysis} --- 
We perform a likelihood-ratio test, which assesses the goodness of fit 
using the following statistic
\begin{equation}
\lambda_{\rm LR}= 2 \ln \frac{\mathcal{L}_{\rm max}(\mathcal{H}_1)}
{\mathcal{L}_{\rm max}(\mathcal{H}_0)}, 
\end{equation}
where $\mathcal{H}_0$ is the {\it null hypothesis} that the timing 
residuals contain only the noise contributions, and $\mathcal{H}_1$ 
is the {\it signal hypothesis} that the DPDM signal is present.
In both hypotheses the noise parameters $\boldsymbol{\vartheta}_{\rm W}$ 
(white noise parameters), $\boldsymbol{\vartheta}_{\rm SN}$ 
(spin noise parameters), $\boldsymbol{\vartheta}_{\rm DM}$ 
(dispersion-measure noise parameters), and $\boldsymbol{\vartheta}_{\rm BN}$ 
(band noise parameters) are fixed at their best-fit values obtained 
from single pulsar analyses. The free parameters include 
$\boldsymbol{\vartheta}_{\rm BE}$ (Bayes ephemeris parameters, vary in both 
$\mathcal{H}_0$ and $\mathcal{H}_1$; see Ref.~\cite{bayes_ephmeris} for 
details) and $\boldsymbol{\vartheta}_{\rm DPDM}$ (DPDM parameters, vary 
only in $\mathcal{H}_1$). The priors of all the parameters mentioned 
above are listed in Supplemental Table~S2 \cite{supp}.

We adopt the Bayesian approach to derive the constraints on the 
DPDM parameters. We set the upper limit on the coupling constant, 
$\bar{\epsilon}$, at 95\% credibility as
\begin{align}
0.95  =  \int_0^{\overline{\epsilon}} P(\epsilon)d\epsilon \nonumber  \int \mathcal{L}(\epsilon,\boldsymbol{\vartheta}'_{\rm DPDM},
\boldsymbol{\vartheta}_{\rm BE}) P(\boldsymbol{\vartheta}'_{\rm DPDM})
P(\boldsymbol{\vartheta}_{\rm BE})\times\\ d\boldsymbol{\vartheta}'_{\rm DPDM}
d\boldsymbol{\vartheta}_{\rm BE},
\end{align}
where $\boldsymbol{\vartheta}'_{\rm DPDM}$ are the DPDM parameters 
excluding $m_A$ and $\epsilon$, and $P$ are the prior probabilities of
those parameters (see Supplemental Table~S2 \cite{supp}). Here the dark 
photon mass $m_A$ is singled out from $\boldsymbol{\vartheta}_{\rm DPDM}$ 
as we search for possible signals for a given range of masses. 
In the computation of the upper limits, we fix both the white noise 
and the red noise parameters as their best-fit values obtained in 
single-pulsar analyses. 

{\it Results and discussion} ---
Our main results about the $95\%$ upper limits on the coupling constant 
$\epsilon^2$, where $\epsilon$ is the dimensionless coupling constant 
$\epsilon$ and $\epsilon e$ is the coupling strength of the $U(1)_B$ or 
$U(1)_{B-L}$ interaction with $e$ being the electromagnetic coupling,
are shown in Fig.~\ref{Fig-bayes-upperlimit}. For comparison, the 
constraints on these parameters from experiments testing the weak 
equivalence principle (WEP), e.g., Ref.~\cite{MICROSCOPE}, are also 
presented. See the Supplemental Material \cite{supp} for a description
of the WEP results. We find that our study imposes significantly 
improved limits in the low-mass region with $m_A\lesssim 3\times
10^{-22}$~eV for $U(1)_{B}$ and $m_A\lesssim 10^{-22}$~eV for $U(1)_{B-L}$. 
Compared with the WEP results, the PPTA constraints on $\epsilon$ 
are stronger by one to two orders of magnitude.

We find that for a few particular frequencies, the signal hypothesis 
fits the PPTA DR2 data better than the null hypothesis. The strongest 
``signal" is at $f\simeq1.02\times10^{-7}~{\rm Hz}$. The corresponding 
$m_A$ and $\epsilon$ are given in Supplemental Table~S3 \cite{supp}. We note that the 
favored value of the coupling constant has already been ruled out by 
existing WEP experiments \cite{MICROSCOPE}. Therefore, it may be due 
to some unmodeled noise effect. A similar spurious signal also appeared 
in the analysis Ref.~\cite{Porayko2018}, which was speculated to be 
caused by the unmodeled perturbations of Solar System barycenter from Mercury. 

In order to have a better understanding of the ``signal" at 
$f\simeq1.02\times10^{-7}~{\rm Hz}$, we reduce the number of pulsars 
and perform a consistency test. We choose two sets of pulsars: The first one includes five pulsars that contribute the most to this strongest “signal”, the second one is the same except that we do not included PSR J0437-4715. Additionally, we allow red-noise parameters 
$\boldsymbol{\vartheta}_{\rm SN}$, $\boldsymbol{\vartheta}_{\rm DM}$, 
and $\boldsymbol{\vartheta}_{\rm BN}$ to vary along with 
$\boldsymbol{\vartheta}_{\rm DPDM}$ and $\boldsymbol{\vartheta}_{\rm BE}$. 
We calculate the Bayes factor between the null and the signal hypothesis. 
We find that the ``signal'' is mainly contributed by PSR J0437-4715. 
For the analysis with the other four pulsars excluding PSR J0437-4715 gives
a logarithmic Bayes factor of 37 in favor of the $\mathcal{H}_1$ hypothesis 
at $f\simeq 6.58\times 10^{-8}~{\rm Hz}$, which is significantly smaller 
than the five-pulsar analysis. The results of the tests are presented in 
Supplemental Table~S4 \cite{supp}. Note also that the best-fit frequency differs 
from that of the five-pulsar analysis, and the best-fit coupling constant
lies above the WEP upper limit, indicating that this may be due to 
unmodelled noise. More careful studies need to be carried out in the future 
to understand the sources of these systematics.

We note that the mass range with the strongest constraint is below the 
typical fuzzy dark matter mass ($10^{-22}$~eV). Currently there is no 
consensus on the constraints on the mass of the fuzzy dark matter. 
The stellar kinematics of dwarf spheroidal galaxies tend to favor a 
lower mass \cite{Gonzalez2017,Kendal2020} (a few times of $10^{-23}$~eV), 
while the halo mass function derived from the Lyman-$\alpha$ forest 
suggests a higher lower limit \cite{Hui2017,Armengaud2017} 
($\gtrsim10^{21}$~eV). Most of these constraints rely on assumptions 
about the structure formation in the fuzzy dark matter scenario. 
More detailed studies of the interplay between the fuzzy dark matter 
and the baryonic matter, such as the dark matter-to-baryon mass ratio 
or various baryonic effects in fuzzy dark matter simulations, are 
required to reach a robust and consistent result.

The gray shaded regions in Fig.~\ref{Fig-bayes-upperlimit} indicate
the parameter space for which the ``gravitational effect'' due to the
space-time metric oscillation induced by the wavelike DPDM field dominates 
over the gauge interaction discussed in this work. See the Supplemental Material \cite{supp}
for more details of the estimate of the gray regions. In such parameter regions,
dedicated analysis with both effects being included in the signal model is
required and will be studied in future work.

{\it Summary} ---
Ultralight fuzzy dark matter is proposed as an attractive candidate
of dark matter in the universe which helps solve the small-scale 
crises of the classical cold dark matter scenario. Using the precise 
timing observations of 26 pulsars by the PPTA project, we study the 
possible couplings between ultralight dark matter and ordinary matter. 
Taking the DPDM as an example, we obtain by far the strongest
constraints on the parameters of the DPDM. The upper limits on the 
dimensionless coupling constant $\epsilon$ derived in our study are 
improved by up to two orders of magnitude when the dark photon mass 
is smaller than $3\times10^{-22}$~eV ($10^{-22}$~eV) for the 
$U(1)_{B}$ ($U(1)_{B-L}$) scenario.

The search sensitivity for the DPDM is expected to improve significantly 
in the near future as more pulsars are monitored with continually extending
data spans by the worldwide PTA campaigns including the 
PPTA, the North American Nanohertz Observatory for Gravitational Waves 
(NANOGrav \cite{NANOGrav}), and the European PTA
(EPTA \cite{EPTA}), which have jointly formed the International PTA (IPTA \cite{IPTA}). The Five-hundred-meter Aperture 
Spherical Telescope (FAST \cite{FAST}) and the Square Kilometer Array 
(SKA \cite{SKA}) are also expected to join the IPTA collaboration.
These efforts are likely to bring revolutionary progress in studying 
a wide range of dark matter models, and more generally in answering 
the related fundamental questions in physics.

{\bf Acknowledgements} 
This work is supported by the National Natural Science Foundation of China 
(No. 11722328, No. 11851305, No. 12003069, No. 11947302, No. 11690022, No. 11851302, No. 11675243,
and No. 11761141011).	
X.X. is supported by Deutsche Forschungsgemeinschaft under Germany's Excellence Strategy EXC2121 ``Quantum Universe'' 390833306.
Q.Y. is supported by the Key Research Program of the Chinese Academy of Sciences (No. XDPB15) 
and the Program for Innovative Talents and Entrepreneur in Jiangsu. 
J.S. is supported by the Strategic Priority Research Program of the 
Chinese Academy of Sciences under Grants No. XDB21010200 and No. XDB23000000.
X.J.Z., A.D.C., B.G., D.J.R. and R.M.S. are supported by ARC CE170100004. 
Z.Q.X. is supported by the Entrepreneurship and Innovation Program of Jiangsu Province.
Y.Z. is supported by U.S. Department of Energy under Award Number DESC0009959.
R.M.S. is the recipient of ARC Future Fellowship FT190100155.
J.W. is supported by the Youth Innovation Promotion Association of Chinese Academy of Sciences.
S.D. is the recipient of an Australian Research Council Discovery Early Career Award (DE210101738) funded by the Australian Government.
The Parkes radio telescope is part of the Australia Telescope, which is funded by the Commonwealth Government for operation as a National Facility managed by CSIRO. 



\setcounter{figure}{0}
\setcounter{table}{0}
\renewcommand\thefigure{S\arabic{figure}}
\renewcommand\thetable{S\arabic{table}}

\section*{Supplemental Material}

\subsection{Noise model}

Following \cite{Porayko2018}, we assume ${\Delta t}_{\rm Noise}$ follows a multivariate Gaussian distribution with a covariance matrix $\boldsymbol{C}^{\rm tot}$. In the most general case, for each pulsar, $\boldsymbol{C^{\rm tot}}$ contains four terms:
\begin{equation}\label{noisemodel}
\boldsymbol{C^{\rm tot}} = \boldsymbol{N} + \boldsymbol{C}^{\rm SN} + 
\boldsymbol{C}^{\rm DM} + \boldsymbol{C}^{\rm BN},
\end{equation}
where $\boldsymbol{N}$ accounts for white noise, $\boldsymbol{C}^{\rm SN}$, $\boldsymbol{C}^{\rm DM}$ and $\boldsymbol{C}^{\rm BN}$ arise from red-noise processes.  $\boldsymbol{C}^{\rm SN}$ is the spin noise that describes the time-correlated spin noise which might be induced by rotational irregularities of the pulsar. $\boldsymbol{C}^{\rm DM}$ accounts for the time variations of the dispersion measures.
$\boldsymbol{C}^{\rm BN}$ is the band noise which is used to account for unknown additional red noise that is only present in a specific radio-frequency band \cite{red_noise_exp_2}.
In our analysis, we only include band noise for two pulsars that are known to exhibit significant band noise \cite{red_noise_exp_2}, PSR J0437-4715 and J1939+2124.
We apply the band noise for these two pulsars in four bands: 10 cm, 20 cm, 40 cm and 50 cm.
All the terms in Eq.~(\ref{noisemodel}) are $n\times n$ matrices, with $n$ being the number of ToAs for each pulsar. 
In practice, the white noise of the actual timing data is usually larger than the ToA measurement uncertainties. Empirically, for each receiver/backend system and each pulsar, a re-scaling factor {\tt EFAC} 
(Error FACtor) is multiplied to the measurement uncertainty, and an extra white noise component {\tt EQUAD} (Error added in QUADrature) is also included \cite{white_1,white_2,white_3}. The rescaled ToA uncertainty $\sigma_s$ is related to the original uncertainty $\sigma$
\begin{equation}\label{noise}
\sigma_s^2 = ({\tt EFAC} \times \sigma)^2 + {\tt EQUAD}^2.
\end{equation}



We model the power spectra of the spin noise, band noise and dispersion-measure noise as power-law functions, 
\begin{equation}
P(f)=\frac{A^2}{12\pi}\left(\frac{f}{\rm yr^{-1}}\right)^{-\gamma}~{\rm yr}^3,
\label{powerlaw}
\end{equation}
where $A$ and $\gamma$ are the noise amplitude and noise spectral index at a reference frequency of ${\rm yr^{-1}}$, respectively.
The noise covariance matrices are,
\begin{eqnarray}
&&\boldsymbol{C}_{ij}^{\rm SN,BN} = \int_{f_L} {\rm d}f P(f) 
\cos\left(2\pi f (t_i-t_j)\right),\label{integration}\\
&&\boldsymbol{C}_{ij}^{\rm DM} = \int_{f_L} {\rm d}f P(f) \left(\frac{\nu^{\rm ref}}{\nu^{\rm obs}_i}\right)^2\left(\frac{\nu^{\rm ref}}{\nu^{\rm obs}_j}\right)^2
\cos\left(2\pi f (t_i-t_j)\right),\label{integration2}
\end{eqnarray}
here $t_i$ and $t_j$ are $i$-th and $j$-th ToAs, $\nu^{\rm obs}_i$ and $\nu^{\rm obs}_j$ are their respective radio frequencies, $f_L$ is the low-frequency cutoff, $\nu^{\rm ref}=1400$~MHz is the reference radio frequency.
Earlier work shows that setting $f_L=1/T$, with $T$ being the observational time interval, is sufficient for pulsar timing analysis due to the fitting for the quadratic spin-down in the timing model \cite{red_noise_1,temponest}.
Note that in Eq.~(\ref{integration}), the elements of band-noise covariance matrics is non-zero only when radio frequencies of $i$-th and $j$-th ToAs belong to the same frequency band.

\begin{figure*}
\centering
\includegraphics[width=\textwidth]{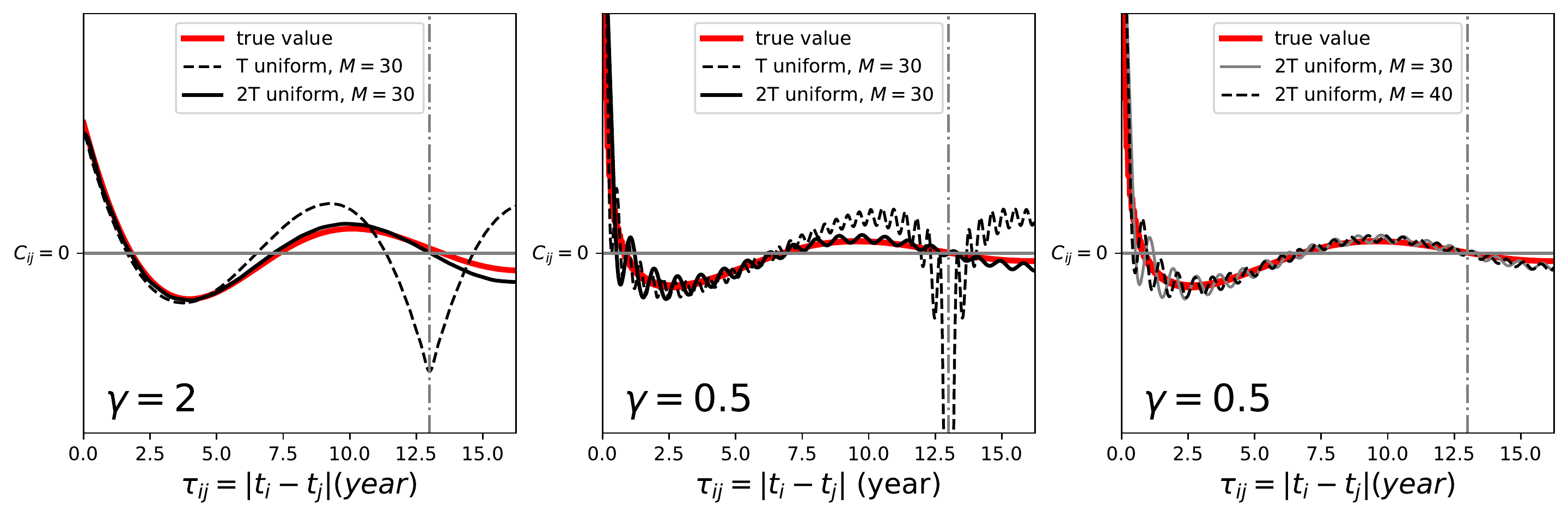}
\caption{The comparison between the 2T uniform approximation and T uniform approximation of Eq.~(\ref{integration}) with different $\gamma$. The y-axis is the value of the noise covariance matrix element at different $\tau_{ij}$. The red thick lines are the true values of Eq.~(\ref{integration}) without disretization and high frequency cut. In first two panels, the solid black lines are the results from the 2T uniform approximation, the dashed black lines are from the T uniform approximation. In the last panel, we show the 2T uniform approximation of Eq.~(\ref{integration}) when the time-frequency modes $M$ equals to $30$ and $40$ with solid gray line and dashed black line respectively. As an illustration, we choose $T\approx13.1$ years, which is shown as the vertical dashed gray line in these plots. The approximation may create additional periodic patterns. To remove the fake signal, we change the number of time-frequency modes from $M=30$ to $M=40$ in the likelihood-ratio test. If the excess only exists in one of the choices of $M$, it will not be identified as our signal.}
\label{Fig-compare-intervals}
\end{figure*}

We adopt the ``time-frequency'' method to calculate the integral in Eq.~(\ref{integration},\ref{integration2}). Namely, we approximate the integral as a discrete summation of different Fourier frequency modes.
In this approximation, a uniform interval of Fourier frequency modes is usually adopted up to an upper limit.
One may choose an interval of ${\rm d}f\equiv \Delta f = 1/T$, which we call ``T uniform".
However, such a choice is sub-optimal \cite{red_noise_2}. Instead, we adopt the ``2T uniform" sampling where $\Delta f = 1/(2T)$ for the approximation of Eq.~(\ref{integration},\ref{integration2}).
We found a better approximation is achieved with the ``2T uniform" interval. On this basis, we find that additional periodic patterns may appear due to the approximation, when $\gamma$ is small. We also compare the influence of the total number of Fourier frequency modes (time-frequency modes) $M$ when $M=30$ and $M=40$, and find that the frequencies of their periodic patterns are different. This is used to remove the potential fake signal introduced during this calculation. The discussions above are demonstrated in Fig.~\ref{Fig-compare-intervals}.

In the ``time-frequency" method, the covariance matrices in Eq.~(\ref{integration}, \ref{integration2}) can be rewritten as
\begin{eqnarray}
\boldsymbol{C} &\approx& \boldsymbol{C}^{(S)} + \boldsymbol{C}^{(C)}
=\boldsymbol{F}^{(S)}\boldsymbol{\Phi}(\boldsymbol{F}^{(S)})^T + \boldsymbol{F}^{(C)}\boldsymbol{\Phi}(\boldsymbol{F}^{(C)})^T,
\end{eqnarray}
with $\boldsymbol{\Phi}_{\alpha\beta}=(A^2/12\pi^2)(1/T)(f_{\alpha})^{-\gamma}\delta_{\alpha\beta}$.
For spin and band noise, $\boldsymbol{F}^{(S)}_{i\alpha}=\sin\left(2\pi f_\alpha t_i\right)$ and $\boldsymbol{F}^{(C)}_{i\alpha}=\cos\left(2\pi f_\alpha t_i\right)$;
For dispersion-measure noise $\boldsymbol{F}^{(S)}_{i\alpha}=(\nu^{\rm ref}/\nu^{\rm obs}_i)^2\sin\left(2\pi f_\alpha t_i\right)$ and $\boldsymbol{F}^{(C)}_{i\alpha}=(\nu^{\rm ref}/\nu^{\rm obs}_i)^2\cos\left(2\pi f_\alpha t_i\right)$.
Here $i$ is the index for ToAs, and $\alpha$ is used to label the discretized Fourier frequency. We follow Ref.~\cite{Porayko2018} in terms of the definition and calculation of the likelihood function.

The parameter description and prior ranges of the noise model is given in Table~\ref{table-parameter}.
In Table~\ref{table-pulsar} we also present the posterior credible intervals for spin noise and dispersion-measure noise based on the results of single pulsar analyses, where we analyze the noise properties of each pulsar independently.
In Bayesian analyses detailed below, we fix the noise parameters at their maximal likelihood values obtained from single pulsar analyses in some cases to reduce the computational costs. They are referred to as the ''best fit'' parameter values, which are available in our code repository.

\subsection{Simulation of the DPDM background}

Here we provide details on how to simulate the DPDM background. 
The interference among the wave-functions of these dark photon particles 
induces fluctuations on the dark matter field strength. Here we study this 
amplitude distribution, and it will be used to inform our priors in the 
statistical analysis. In addition, by studying the spatial correlations 
among various quantities of the dark photon background, we demonstrate 
the concept of coherence length. Within the coherence length, the DPDM
can be approximated as a plane wave, whereas this approximation breaks 
down once the spatial separation becomes larger.

The dark photon background $\vecA^{(tot)}$ can be modeled as the summation 
of many, $N\gg 1$, freely propagating planewaves,
\begin{equation}
\vecA^{(tot)}(t,\vecx)
=\sum_{n=1}^N \vecA^{(n)}(t,\vecx)
=\sum_{n=1}^N \vecA_{0}^{(n)}\sin(\omega^{(n)} t -\boldsymbol{k}^{(n)}\cdot \vecx + \alpha^{(n)}),
\end{equation}
where $n$ is the index for each dark photon particle and $\alpha^{(n)}$ 
is the phase term in a plane wave.  $\vecA_{0}^{(n)}$ is the gauge 
potential vector, whose magnitude is the same for all particles, but 
the direction is isotropically distributed. $\omega^{(n)}$ and 
$\boldsymbol{k}^{(n)}$ are the particle's energy and momentum, 
distributed according to Boltzmann velocity distribution with virial 
velocity $v_{vir}\sim 220$ km/s.  

The normalization of the dark photon field is determined by the local 
dark matter energy density $\rho_{0}$,
\begin{equation}\label{loc_density}
\int_{0}^{T} {\rm d}t\int_{V} {\rm d}^3\vecx~\frac{1}{2} m_A^2 |\vecA^{(tot)}(t,\vecx)|^2 =(VT)~  \rho_{0},
\end{equation}
where $V$ and $T$ need to be sufficiently large, much larger than the 
coherence volume and time. Since $v_{vir}\ll c$, we have 
$\omega^{(n)}\simeq m_A$. For the mass range that we are interested in, 
the coherence time is much larger than the observation time. Also as 
discussed previously, the phase variations, induced by the motions of 
pulsars and the Earth in the Galaxy during the period of observation, 
are negligible. Thus in our analysis, we can approximate the wave-function 
for the DPDM as
\begin{equation}
A_{i}^{(tot)}(\vecx)
=A_{0,i}^{(tot)}(\vecx)\sin(m_A t + \alpha_i(\vecx)),
\end{equation}
where the index $i$ labels the spatial component of the dark photon field, 
i.e. $\{x,y,z\}$. The averaged magnitude of the gauge potential component 
can written as	
\begin{equation}\label{ExpA}
E [(A^{(tot)}_{0,i})^2] = \frac{2}{3}\frac{\rho_{0}}{m_A^2}.
\end{equation}	
For simplicity we introduce a normalized DPDM amplitude
\begin{equation}
\tilde{A}^{(\rm{tot})}_{0,i} \equiv  \frac{\sqrt{3}m_A}{\sqrt{2\rho_{0}}}A^{(\rm{tot})}_{0,i}.
\end{equation}

\begin{figure*}
\centering
\includegraphics[width=0.45\textwidth]{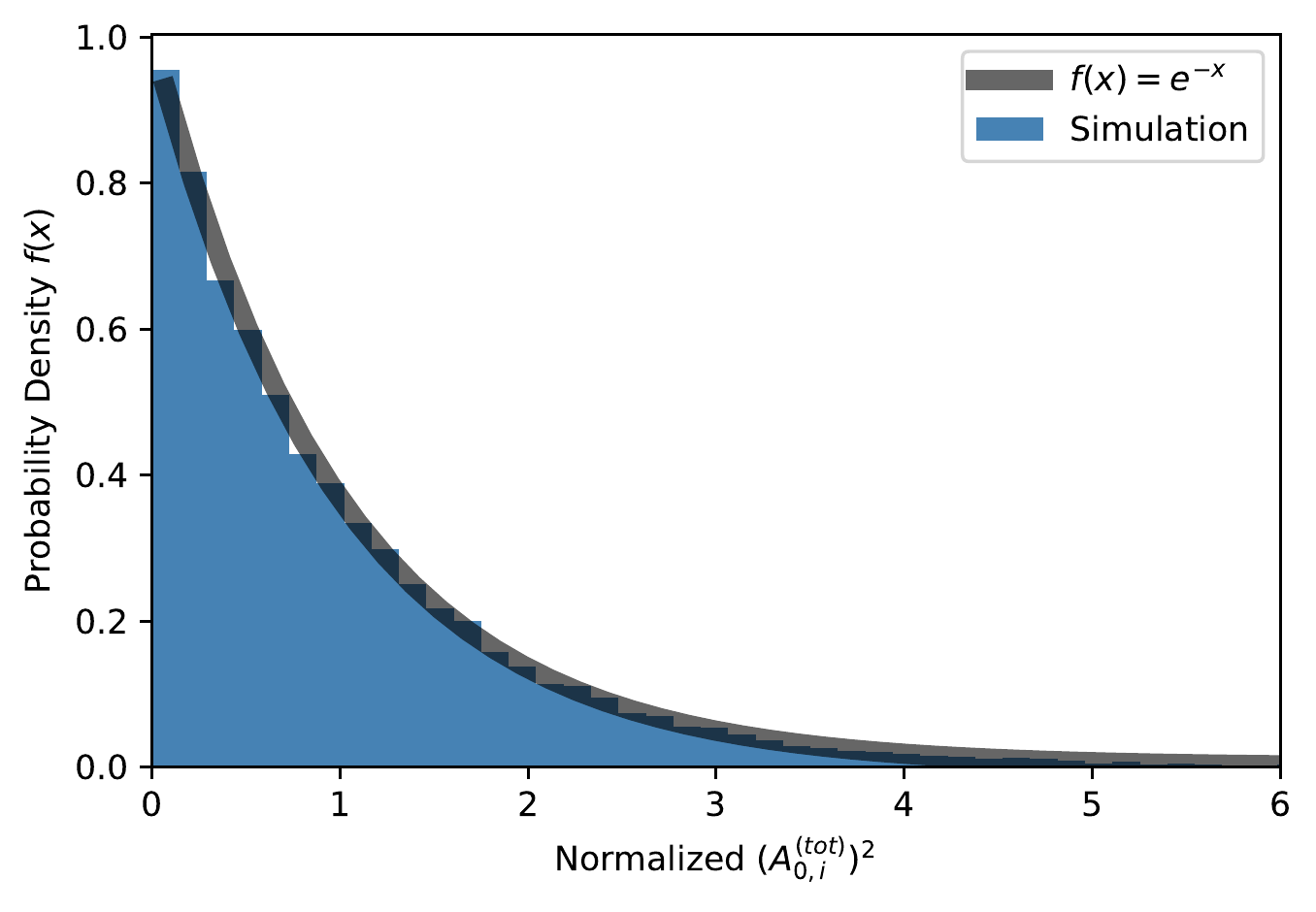}
\includegraphics[width=0.51\textwidth]{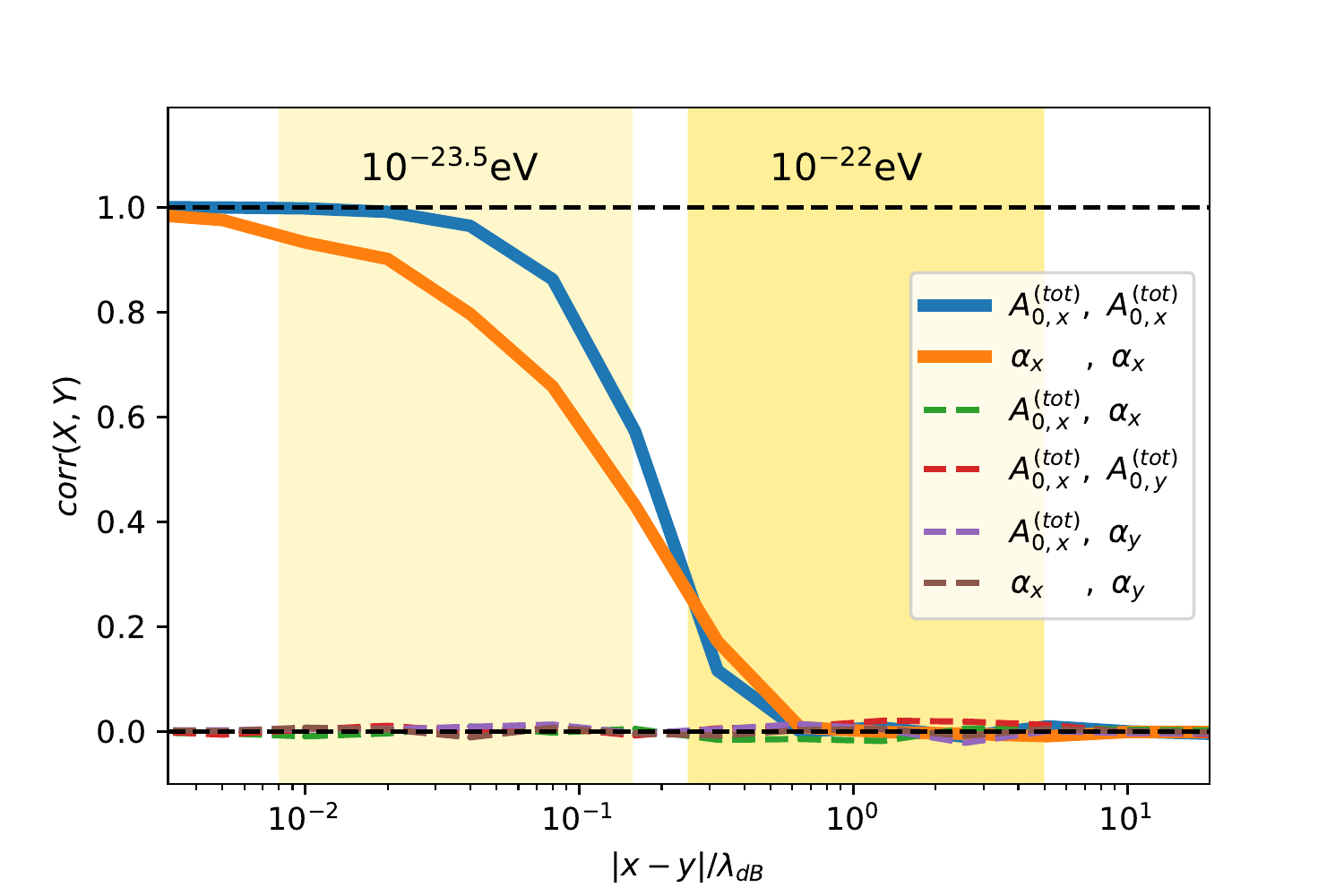}
\caption{Left: the probability distribution of normalized $A^{(tot)}_{0,i}$ ($\tilde{A}^{(tot)}_{0,i}$)
from the simulation. The black curve shows an exponential function fit. 
Right: the correlations between different variables as functions of spatial 
separation. $\lambda_{dB}=2\pi/(m_A v_{vir})$ is the de Broglie wavelength. 
The correlation is non-zero only when the random variables $X$ and $Y$ 
have the same spatial indices, and thus the three spatial components 
of the DPDM field are always independent. The two yellow bands indicate 
the distances between the Earth and most pulsars ($0.1-2$ kpc) when 
$m_A=10^{-23.5}$~eV and $m_A=10^{-22}$~eV. 
}
\label{Fig-Exp-fit}
\end{figure*}

To study the statistic behaviors of the DPDM, we set 
$N=2^{15}$ and simulate the background field for $10^5$ times. This ensures 
the results we obtain are statistically stable. We first focus on the DPDM
amplitude. In the left panel of Fig.~\ref{Fig-Exp-fit}, we show the 
distribution of $(\tilde{A}^{(\rm{tot})}_{0,x})^2$ at $\vecx = \boldsymbol{0}$. 
This distribution can be well fit by an exponential function, which will 
be used as the prior probability of $(\tilde{A}^{(\rm{tot})}_{0,i})^2$ 
in our Bayesian analysis.

Next we study the coherence of the dark photon field as a function of 
the spatial separation between two points. The coherence between two 
quantities, $X$ and $Y$, can be described by their two-point correlation 
function as
\begin{equation}
corr(X,Y)=\frac{E[(X-\mu_{X})(Y-\mu_Y)]}{\sigma_X\sigma_Y},
\end{equation}
where $\mu_{X}$, $\mu_{Y}$ are expectation values, and $\sigma_X$, 
$\sigma_Y$ are standard deviations. Let us consider 
$X\in\{ \tilde{A}^{(\rm{tot})}_{0,i}(0,0,0),~ \alpha_i(0,0,0) \}$ and 
$Y\in\{ \tilde{A}^{(\rm{tot})}_{0,i}(d,0,0),~ \alpha_i(d,0,0) \}$, 
where $d$ represents the separation between two points. In the right
panel of Fig.~\ref{Fig-Exp-fit}, we show $corr(X,Y)$ as a function of 
spatial separation. Here we clearly see that the correlation function 
reduces significantly once the spatial separation is longer than the 
coherence length, i.e., the de Broglie wavelength. We compare our 
result with the separations among the Earth and pulsars. We find that 
when $m_A>10^{-22}$ eV, it is safe to assume that the dark photon 
field at these locations are completely uncorrelated. 

\subsection{Limits from the WEP experiment}
The experiments testing the weak equivalence principle (WEP) can also impose stringent constraints on the coupling constant of the $U(1)_B$ or $U(1)_{B-L}$ gauge force. In the WEP experiments, the dark photon is not required to be the dark matter. Here we compare our results with those obtained by WEP experiments. For the mass regime that we are interested in this study, the best limit is given by the satellite experiment, MICROSCOPE \cite{MICROSCOPE}. The effect is characterized by the E\"otv\"os parameter as
\begin{equation}
\eta_{\rm E-s} \simeq\frac{e^2 \epsilon^2}{4\pi G}\left[\left(\frac{q_1}{m_1}\right)
-\left(\frac{q_2}{m_2}\right)\right]\frac{q_E}{m_E},
\end{equation}
where $q_{1,2,E}$ and $m_{1,2,E}$ are the ``dark charges'' and masses of the two test bodies in the MICROSCOPE satellite and the Earth, respectively. The best sensitivity achieved by the MICROSCOPE  experiment \cite{MICROSCOPE} is $|\eta|<2.7\times10^{-14}$ at $2\sigma$ level. This can be translated to constraints on coupling constants as $\epsilon>10^{-23.28}$ for $U(1)_B$ and $\epsilon>10^{-24.07}$ for $U(1)_{B-L}$. The results are consistent with Ref.~\cite{Pierre1}. In Fig.~\ref{Fig-bayes-upperlimit}, we find our results improves upon that of MICROSCOPE when $m_A\lesssim 3\times10^{-22}$~eV for $U(1)_{B}$ and $m_A\lesssim 10^{-22}$~eV for $U(1)_{B-L}$, by up to two orders of magnitudes.

\subsection{Gravitational effect of DPDM on the ToA}
The quantum pressure of the ultralight bosonic dark matter field can induce a periodic oscillation of the metric. This will affect the propagation of photons and lead to detectable oscillatory effects at pulsar timing experiments. Such an effect has been studied in Refs.~\cite{gw_effect_1,Porayko2018} in the case of ultralight scalar field. Ultralight vector field, such as the dark photon, can also be searched in this way. Thus it is instructive to compare this pure gravitational effect with the effect studied in this paper.

According to Refs.~\cite{gw_effect_1,gw_effect_2}, the timing residual caused by the oscillating quantum pressure can be written as,
\begin{eqnarray}
&&\Delta t_{\rm scalar} =\frac{\Psi_c}{ \pi f} \times\sin (\Delta\alpha_p) \cos(2\pi f t +2 \alpha_e +\Delta\alpha_p),\\
&&\Delta t_{\rm vector} \leq 3\times \Delta t_{\rm scalar}.
\end{eqnarray}
Here $\Psi_c\equiv G\rho_{0}/(\pi f^2)$ with $f = m_{\rm DM}/\pi$, and $m_{\rm DM}$ is the mass of the dark matter particle. The root-mean-square value of $\Delta t$ is
\begin{eqnarray}
&&\sqrt{\langle\Delta t_{\rm scalar}^2\rangle} \simeq 0.5 ~ \pi\times \frac{G\rho_{0}}{ m_{\rm DM}^3},\\
&&\sqrt{\langle\Delta t_{\rm vector}^2\rangle} \simeq 1.5 ~ \pi\times \frac{G\rho_{0}}{ m_{\rm DM}^3},
\end{eqnarray}
where we take the maximal value of vector dark matter's timing residual. Meanwhile, the effect induced by the gauge coupling of the DPDM has the root-mean-square value as
\begin{eqnarray}
&&\sqrt{\langle(\Delta t_{\rm dp}^{(B)})^2\rangle}\simeq 0.577\ \epsilon e  \frac{ \sqrt{2\rho_{0}}}{m_A^2},\\
&&\sqrt{\langle(\Delta t_{\rm dp}^{(B-L)})^2\rangle}\simeq 0.456\ \epsilon e  \frac{ \sqrt{2\rho_{0}}}{m_A^2}.
\end{eqnarray}
where the prefactors are obtained through our numerical simulation. By comparing the root-mean-square value of these two effects, we find that the gauge interaction of the dark photon induces a stronger effect when $\epsilon\times m_{A}/{\rm eV}>10^{-48.35}$ for $U(1)_B$ and $10^{-48.24}$ for $U(1)_{B-L}$. This comparison is shown as the black dashed line in Fig.~\ref{Fig-bayes-upperlimit}.

\begin{table*}
\caption{Properties of the 26 pulsars used in this analysis. This table doesn't include the band noise properties of pulsars J0437-4715 and J1939+2134.}
\label{table-pulsar}	
\begin{tabularx}{\textwidth}{XXXXXXXX}			
\hline\hline
Pulsars & $N_{\rm obs}$ & $T$~(years) & $\bar{\sigma}\times 10^{-6}$~(s) & $\log A_{\rm SN}$ & $\gamma_{\rm SN}$ & $\log A_{\rm DM}$ & $\gamma_{\rm DM}$\\
\hline
\textbf{J0437-4715}&29262&15.03& 0.296 &$-15.76^{+0.17}_{-0.18}$&$6.63^{+0.17}_{-0.13}$&$-13.05^{+0.10}_{-0.08}$&$2.26^{+0.32}_{-0.44}$\\
J0613-0200&5920&14.20&2.504&$-14.63^{+0.77}_{-0.68}$&$4.93^{+1.33}_{-1.61}$&$-13.02^{+0.08}_{-0.08}$&$0.95^{+0.33}_{-0.31}$\\
J0711-6830&5547&14.21&6.197&$-12.85^{+0.14}_{-0.16}$&$0.97^{+0.64}_{-0.55}$&$-14.54^{+0.72}_{-0.89}$&$4.43^{+1.68}_{-1.72}$\\
J1017-7156&4053&7.77&1.577&$-12.89^{+0.07}_{-0.07}$&$0.54^{+0.53}_{-0.37}$&$-12.72^{+0.06}_{-0.06}$&$2.18^{+0.45}_{-0.44}$\\
J1022+1001&7656&14.20&5.514&$-12.79^{+0.12}_{-0.13}$&$0.54^{+0.55}_{-0.37}$&$-13.04^{+0.10}_{-0.12}$&$0.58^{+0.47}_{-0.36}$\\
J1024-0719&2643&14.09&4.361&$-14.28^{+0.27}_{-0.20}$&$6.51^{+0.35}_{-0.60}$&$-14.53^{+0.54}_{-0.56}$&$5.22^{+1.14}_{-1.18}$\\
J1045-4509&5611&14.15&9.186&$-12.75^{+0.24}_{-0.40}$&$1.58^{+1.28}_{-0.93}$&$-12.18^{+0.09}_{-0.08}$&$1.86^{+0.36}_{-0.32}$\\
J1125-6014&1407&12.34&1.981&$-12.64^{+0.11}_{-0.12}$&$0.51^{+0.55}_{-0.37}$&$-13.14^{+0.19}_{-0.21}$&$3.36^{+0.73}_{-0.66}$\\
J1446-4701&508&7.36&2.200&$-16.46^{+2.88}_{-3.17}$&$2.74^{+2.49}_{-1.89}$&$-13.49^{+0.32}_{-1.87}$&$2.48^{+1.92}_{-1.45}$\\
J1545-4550&1634&6.97&2.249&$-17.33^{+2.50}_{-2.55}$&$3.25^{+2.45}_{-2.18}$&$-13.40^{+0.24}_{-0.38}$&$3.90^{+1.61}_{-1.09}$\\		
J1600-3053&7047&14.21&2.216&$-17.63^{+2.10}_{-2.29}$&$3.28^{+2.34}_{-2.15}$&$-13.27^{+0.12}_{-0.13}$&$2.79^{+0.43}_{-0.40}$\\
J1603-7202&5347&14.21&4.947&$-12.82^{+0.14}_{-0.16}$&$1.01^{+0.67}_{-0.60}$&$-12.66^{+0.10}_{-0.09}$&$1.44^{+0.40}_{-0.38}$\\
J1643-1224&5941&14.21&4.039&$-12.32^{+0.08}_{-0.09}$&$0.51^{+0.42}_{-0.34}$&$-12.27^{+0.07}_{-0.07}$&$0.55^{+0.32}_{-0.29}$\\	
J1713+0747&7804&14.21&1.601&$-14.09^{+0.25}_{-0.38}$&$2.98^{+1.00}_{-0.64}$&$-13.35^{+0.08}_{-0.08}$&$0.53^{+0.32}_{-0.31}$\\
J1730-2304&4549&14.21&5.657&$-17.39^{+2.39}_{-2.51}$&$3.05^{+2.59}_{-2.12}$&$-14.11^{+0.40}_{-0.57}$&$4.22^{+1.42}_{-1.04}$\\
J1732-5049&807&7.23&7.031&$-16.51^{+3.04}_{-2.97}$&$3.29^{+2.37}_{-2.97}$&$-13.38^{+0.54}_{-0.84}$&$4.07^{+1.96}_{-1.93}$\\	
J1744-1134&6717&14.21&2.251&$-13.39^{+0.14}_{-0.15}$&$1.49^{+0.66}_{-0.57}$&$-13.35^{+0.09}_{-0.09}$&$0.86^{+0.40}_{-0.33}$\\
J1824-2452A&2626&13.80&2.190&$-12.56^{+0.13}_{-0.12}$&$3.61^{+0.41}_{-0.39}$&$-12.18^{+0.11}_{-0.10}$&$1.64^{+0.46}_{-0.59}$\\
J1832-0836&326&5.40&1.430&$-16.47^{+2.63}_{-3.09}$&$3.66^{+2.33}_{-2.52}$&$-13.07^{+0.24}_{-0.63}$&$3.77^{+2.00}_{-1.05}$\\
J1857+0943&3840&14.21&5.564&$-14.76^{+0.74}_{-0.50}$&$5.75^{+0.91}_{-1.53}$&$-13.40^{+0.20}_{-0.25}$&$2.66^{+0.83}_{-0.67}$\\
J1909-3744&14627&14.21&0.672&$-13.60^{+0.13}_{-0.12}$&$1.60^{+0.43}_{-0.46}$&$-13.48^{+0.09}_{-0.08}$&$0.69^{+0.38}_{-0.35}$\\
\textbf{J1939+2134}&4941&14.09&0.468&$-14.38^{+0.22}_{-0.18}$&$6.24^{+0.49}_{-0.62}$&$-11.59^{+0.07}_{-0.07}$&$0.13^{+0.19}_{-0.10}$\\
J2124-3358&4941&14.21&8.863&$-14.79^{+0.82}_{-0.67}$&$5.07^{+1.37}_{-1.97}$&$-13.35^{+0.18}_{-0.33}$&$0.95^{+1.11}_{-0.66}$\\
J2129-5721&2879&13.88&3.496&$-15.48^{+1.92}_{-3.54}$&$2.91^{+2.29}_{-1.83}$&$-13.31^{+0.13}_{-0.14}$&$1.07^{+0.65}_{-0.65}$\\
J2145-0750&6867&14.09&5.086&$-12.82^{+0.10}_{-0.11}$&$0.62^{+0.50}_{-0.40}$&$-13.33^{+0.14}_{-0.16}$&$1.38^{+0.54}_{-0.55}$\\
J2241-5236&5224&8.20&0.830&$-13.40^{+0.09}_{-0.08}$&$0.44^{+0.40}_{-0.30}$&$-13.79^{+0.10}_{-0.10}$&$1.42^{+0.61}_{-0.59}$\\
\hline\hline
\end{tabularx}
\end{table*}

\begin{table*}
\centering
\caption{The model parameters of the analysis. $\tilde{A}_0^{p}$ and $\tilde{A}_0^{e,i}$ are normalized amplitudes of the DPDM field. We consider only one set of DPDM parameters for each pulsar, $\tilde{A}_0^{p}$ and $\alpha_p$, while all three spatial components for the Earth are included. This is because only the DPDM oscillation along the line of sight is detectable. * suggests that the parameters only exist in the uncorrelated case.}
\label{table-parameter}
\begin{tabularx}{\textwidth}{XXXX}
\hline\hline
& Parameter & Prior & Description \\
\hline
\multirow{2}*{\shortstack{White Noise $\boldsymbol{\vartheta}_{\rm W}$} }
&{\tt EFAC}  & U[$0.01$,$10$] & one per backend\\
&{\tt EQUAD} & log-U [$10^{-10}$,$10^{-4}$] & one per backend\\
\hline
\multirow{2}*{\shortstack{Spin Noise $\boldsymbol{\vartheta}_{SN}$} }
&$\gamma_{\rm SN}$ & U[$0$,$7$] & one per pulsar\\
&$A_{\rm SN}$ & log-U[$10^{-21}$,$10^{-9}$] & one per pulsar\\
\hline
\multirow{2}*{\shortstack{DM Noise $\boldsymbol{\vartheta}_{\rm DM}$}}
&$\gamma_{\rm DM}$ & U[$0$,$7$] & one per pulsar\\
&$A_{\rm DM}$ & log-U[$10^{-21}$,$10^{-9}$] & one per pulsar \\
\hline
\multirow{2}*{\shortstack{Band Noise $\boldsymbol{\vartheta}_{\rm BN}$\\for J0437 and J1939 }}
&$\gamma_{\rm BN}$ & U[$0$,$7$] & one per band\\
&$A_{\rm BN}$ & log-U[$10^{-21}$,$10^{-9}$] & one per band \\

\hline
\multirow{6}*{\shortstack{Dark photon \\Parameters $\boldsymbol{\vartheta}_{\rm DPDM}$}}
&$\alpha_p$ & U[$0$,$2\pi$] & one per pulsar\\
&$\alpha_e^i$  & U[$0$,$2\pi$] & three per PTA\\
&$(\tilde{A}_0^{p})^2$&$f(x)=e^{-x}$&one per pulsar*\\
&$(\tilde{A}_0^{e,i})^2$&$f(x)=e^{-x}$& three per PTA\\\
&$\epsilon$  & log-U[$10^{-28}$,$10^{-16}$] & one per PTA\\
&$m_A$ & log-U[$10^{-23.5}$,$10^{-21}$]& one per PTA\\
\hline

\multirow{6}*{\shortstack{{\tt BayesEphem} \\Parameters $\boldsymbol{\vartheta}_{\rm BE}$}}
&$z_{\rm drift}$ &U[$-10^{-9}$,$10^{-9}$]& one per PTA \\
&$\Delta M_{\rm Jupiter}$&N$(0,1.5\times10^{-11})$& one per PTA\\
&$\Delta M_{\rm Saturn}$&N$(0,8.2\times10^{-12})$& one per PTA\\
&$\Delta M_{\rm Uranus}$&N$(0,5.7\times10^{-11})$& one per PTA\\
&$\Delta M_{\rm Neptune}$&N$(0,7.9\times10^{-11})$& one per PTA\\
&${\rm PCA}_i$&U[$-0.05$,$0.05$]& six per PTA\\
\hline\hline
\end{tabularx}
\end{table*}

\begin{table*}
\caption{The best-fit dark photon parameters that favor the $\mathcal{H}_1$ hypothesis with a total of 26 pulsars.}
\label{GLRT}
\begin{tabularx}{\textwidth}{XXXXX}
\hline\hline
Interaction & $m_A{\rm (eV)}$ & $f{ \rm (Hz)}$ & $\epsilon$ & $\lambda_{LR}^{max}$\\
\hline
$U(1)_B$&$10^{-21.37}$&$1.02\times 10^{-7}$&$10^{-22.68}$&472\\
$U(1)_{B-L}$&$10^{-21.37}$&$1.02\times 10^{-7}$&$10^{-22.87}$&498\\
\hline
\end{tabularx}
\end{table*}

\begin{table*}[t]

\caption{The Bayes factor and maximal-likelihood parameters with different choices of pulsars and DPDM parameters. As a demonstration, here we only show the results for the $U(1)_B$ model with uncorrelated dark photon background field. The ``5 pulsars'' includes J0437-4715, J1022+1001, J1744-1134, J1909-3744 and J2145-0750, while the ``4 pulsars'' excludes J0437-4715 in the analysis.}
\label{BayesFactor}

\begin{tabularx}{\textwidth}{XXXcX}

\hline\hline
\multirow{2}*{Target}
&\multirow{2}*{Range of $m_A{\rm(eV)}$}
&\multirow{2}*{ln(Bayes Factor)}
&Best Fit Parameters
&\multirow{2}*{$\lambda_{LR}^{max}$}\\
&&&$(m_A$,$\epsilon,f)$&\\
\hline
\multirow{3}*{5 pulsars}
&$[10^{-23.5},10^{-22}]$&3.5&$(10^{-22.98}{\rm eV},10^{-24.60},2.51\times10^{-9}{\rm Hz})$&7.2\\
&$[10^{-22},10^{-21}]$&207.0&\multirow{2}*{$(10^{-21.37}{\rm eV},10^{-22.66},1.02\times10^{-7}{\rm Hz})$}&\multirow{2}*{458}\\
&$[10^{-23.5},10^{-21}]$&212.3&& \\
\hline
4 pulsars
&$[10^{-23.5},10^{-21}]$&37.2&$(10^{-21.56}{\rm eV},10^{-22.89},6.58\times10^{-8}{\rm Hz})$&114\\
\hline
\end{tabularx}

\end{table*}



\begin{thebibliography}{99}
\bibitem{fn1}
In order to give a mass to the dark photon, a Higgs boson in the dark section may be introduced. Since such a Higgs is also very weakly coupled to the Standard Model sector, its existence is challenging to be tested. Furthermore, if one pushes the dark Higgs coupling to be very small, takes its vacuum expectation value to be very large, meanwhile maintains the product of these two parameters to be finite, we reach the Stueckelberg limit where the dark Higgs is very heavy and effectively decoupled from the rest of the theory, except the dark photon mass it generates.
\bibitem{fn7}
The occupation number is $\rho_{\rm 0}/m_A\sim 
10^{30}{\rm cm}^{-3}$, where $\rho_{\rm 0}$ is the local dark matter 
energy density.
\bibitem{fn2}
Recently, the NANOGrav team claimed a detection of a common-spectrum stochastic process in their 12.5-yr data set \cite{NANOGrav12.5}. The NANOGrav ``feature'' which is modeled as a power-law spectrum in the low-frequency range, is different from the monochromatic signal induced by the DPDM studied in this work.
\end{thebibliography}
\end{document}